\documentclass[aps,prb,superscriptaddress,twocolumn,nopacs,letter,rpreprint]{revtex4-1}
\usepackage{graphicx,bm,times}
\usepackage{amsmath}
\usepackage{amsfonts}
\usepackage{amssymb}
\usepackage{color}
\usepackage{hyperref}
\hypersetup{
	colorlinks = true,
	allcolors = {blue}
}
\usepackage[utf8]{inputenc}
\usepackage[T1]{fontenc}
\usepackage{graphicx}
\usepackage{amsmath,amssymb}
\usepackage{lineno}
\begin{document}
	
\title{Direct observation of the energy gain underpinning ferromagnetic superexchange in the electronic structure of CrGeTe$\boldsymbol{_3}$}
	
\author{Matthew D. Watson}
\email{matthew.watson@diamond.ac.uk}
\affiliation {SUPA, School of Physics and Astronomy, University of St Andrews, St Andrews KY16 9SS, United Kingdom}

\author{Igor Markovi{\'c}}
\affiliation {SUPA, School of Physics and Astronomy, University of St Andrews, St Andrews KY16 9SS, United Kingdom}
\affiliation{Max Planck Institute for Chemical Physics of Solids, N\"{o}thnitzer Strasse 40, 01187 Dresden, Germay}

\author{Federico Mazzola}
\affiliation {SUPA, School of Physics and Astronomy, University of St Andrews, St Andrews KY16 9SS, United Kingdom}

\author{Akhil Rajan}
\affiliation {SUPA, School of Physics and Astronomy, University of St Andrews, St Andrews KY16 9SS, United Kingdom}

\author{Edgar A. Morales}
\affiliation {SUPA, School of Physics and Astronomy, University of St Andrews, St Andrews KY16 9SS, United Kingdom}
\affiliation{Max Planck Institute for Chemical Physics of Solids, N\"{o}thnitzer Strasse 40, 01187 Dresden, Germay}

\author{David M. Burn}
\affiliation{Magnetic Spectroscopy Group, Diamond Light Source, Didcot, OX11~0DE, United Kingdom}

\author{Thorsten Hesjedal}
\affiliation{Department of Physics, Clarendon Laboratory, University of Oxford, Oxford, OX1~3PU, United Kingdom}

\author{Gerrit \surname{van der Laan}}
\affiliation{Magnetic Spectroscopy Group, Diamond Light Source, Didcot, OX11~0DE, United Kingdom}

\author{Saumya Mukherjee}
\affiliation {Diamond Light Source, Harwell Campus, Didcot, OX11 0DE, United Kingdom}

\author{Timur K. Kim}
\affiliation {Diamond Light Source, Harwell Campus, Didcot, OX11 0DE, United Kingdom}

\author{Chiara Bigi}
\affiliation{Istituto Officina dei Materiali (IOM)-CNR, Laboratorio TASC, Area Science Park, S.S.14, Km 163.5, 34149 Trieste, Italy}
\affiliation{Department of Physics, University of Milano, 20133 Milano, Italy}
\author{Ivana Vobornik}
\affiliation{Istituto Officina dei Materiali (IOM)-CNR, Laboratorio TASC, Area Science Park, S.S.14, Km 163.5, 34149 Trieste, Italy}

\author{Monica  \surname{Ciomaga Hatnean}}
\author{Geetha Balakrishnan}
\affiliation {Department of Physics, University of Warwick, Coventry, CV4 7AL, United Kingdom}
	
\author{Philip D. C. King}
\email{philip.king@st-andrews.ac.uk}
\affiliation {SUPA, School of Physics and Astronomy, University of St Andrews, St Andrews KY16 9SS, United Kingdom}

\date{\today}

\begin{abstract}
We investigate the temperature-dependent electronic structure of the van der Waals ferromagnet, CrGeTe$_3$. Using angle-resolved photoemission spectroscopy, we identify atomic- and orbital-specific band shifts upon cooling through ${T_\mathrm{C}}$. From these, together with x-ray absorption spectroscopy and x-ray magnetic circular dichroism measurements, we identify the states created by a covalent bond between the Te ${5p}$ and the Cr ${e_g}$ orbitals as the primary driver of the ferromagnetic ordering in this system, while it is the Cr ${t_{2g}}$ states that carry the majority of the spin moment. The ${t_{2g}}$ states furthermore exhibit a marked bandwidth increase and a remarkable lifetime enhancement upon entering the  ordered phase, pointing to a delicate interplay between localized and itinerant states in this family of layered ferromagnets.
\end{abstract}
\maketitle

\section{Introduction}

The recent discovery of long-range ferromagnetic order in single/bilayer transition metal chalcogenides and halides~\cite{Huang2017Nature,Gong2017Nature} has opened new opportunities for studying fundamental questions in magnetism,~\cite{Burch2018} and has provided a powerful and highly-tunable platform for the engineering of magnetic interactions~\cite{Burch2018,Li2019AdvMat} and for the creation of novel device structures~ \cite{Wang2019SciAdvFGT,Gibertini2019Review,GongZhang2019Science}. Typically, such two-dimensional magnets are realised by isolating individual layers of magnetic van der Waals materials systems. Even in the parent bulk compounds, however, the microscopic origins and nature of their magnetic ordering has proved controversial, and key quantities such as how the underlying electronic structure evolves into the magnetic state remains completely unexplored to date. In this work, we show how such electronic structure information, extracted from temperature-dependent angle-resolved photoemission spectroscopy (ARPES), can be used to provide a direct probe of atomic- and orbital-specific energy gains underpinning the magnetic ordering.  

\begin{figure*}
	\centering
	\includegraphics[width=0.8\textwidth]{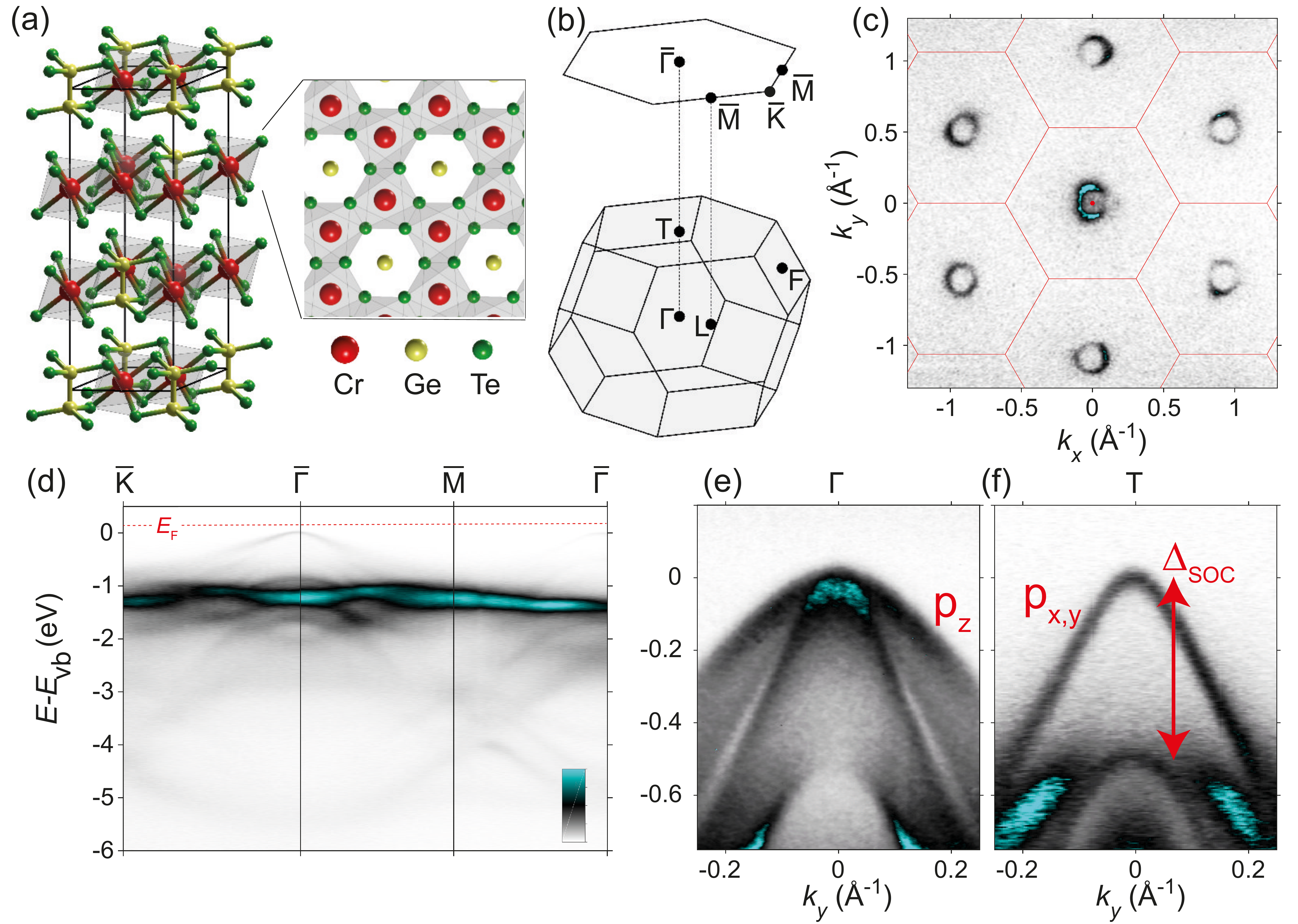}
	\caption{{Paramagnetic electronic structure of CrGeTe$_3$.} a) Crystal structure of CrGeTe$_3$, with trilayer stacking in the unit cell, and b) corresponding 3D Brillouin zone. High-symmetry points are shown, as well as their projection into the hexagonal surface Brillouin zone (barred notation). c) ARPES intensity map ($h\nu=80$~eV) at an energy 0.1 eV below the top of the valence band, showing slightly trigonally-warped contours derived from $p_{x/y}$ states located at each $\bar{\Gamma}$ point. d) ARPES dispersion measured ($h\nu=80$~eV) along high-symmetry directions of the surface Brillouin zone, over an extended binding energy range. A pronounced state with only weak dispersion is evident in addition to the strongly dispersive states. The energy scale is referenced to the valence band maximum, $E_{\mathrm{vb}}$; the chemical potential lies within the band gap, and for the sample shown here is located 0.165 eV above the valence band top (dashed red line). e,f)  Detailed valence band dispersions at high-symmetry points of the 3D Brillouin zone (e: $\Gamma$; $h\nu=22$~eV, f: T; $h\nu=61$ eV). The $p_z$ band disperses significantly along $k_z$ such that it is at higher binding energies than the range shown in f) for the T-point. All data are measured in the paramagnetic phase at $T=100$~K.}
	\label{fig1}
\end{figure*}

We focus here on CrGeTe$_3$, a ferromagnetic semiconductor with a $T_\mathrm{C}$ of $\approx\!63$~K in its bulk form~\cite{Carteaux1995,Ji2013JAPCava,Liu2017PRB}. The transition temperature decreases with reducing thickness, but long-range ferromagnetic order persists down to the bilayer (\textit{i.e.}, when the material is thinned such that it hosts only two Cr-containing layers stacked along the out-of-plane direction) \cite{Gong2017Nature}. Its crystal structure (Fig.~\ref{fig1}(a)) can be considered as a variation on the `CdI$_2$'-type layered 1T transition metal-dichalcogenides\cite{Siberchicot1996}. Compared to meta-stable 1T-CrTe$_2$ \cite{Freitas2015JPhysCondMat}, however, one third of the Cr sites are replaced by Ge dimers. Within a simple ionic picture, the remaining Cr ions are thus in a more stable $3+$ nominal charge state, and are arranged in a honeycomb lattice, similar to the CrX$_3$ trihalide family \cite{McGuire2017CrystalsReview}.

\section{Methods}

Single crystals of Cr$_2$Ge$_2$Te$_6$ were grown using Te as a flux following previously reported procedures~\cite{Ji2013JAPCava}. The crystals were separated from the flux by centrifuging at 500$^\circ$C and the resulting crystal platelets were cleaned free of any excess Te. The crystals were found to exhibit a sharp magnetic transition at $\sim$65 K (see Fig.~SM1 in the Supplemental Material, S.M., \cite{SM}) in good agreement with previous literature reports~\cite{Liu2017PRB}. ARPES measurements were performed at beamlines APE-LE at Elletra, Italy and I05 at Diamond Light Source, UK. Single crystal samples were cleaved in-situ, and ARPES measurements were performed using $p$-polarised photons (photon energies specified in the figure captions) at sample temperatures between $\approx$25 and 150~K. 

Since CrGeTe$_3$ is a semiconductor with a resistivity that becomes large at low temperatures \cite{Ji2013JAPCava}, charging of the sample due to the emission of photoelectrons can becomes a restrictive issue. In practice, we found that certain regions of the sample suffered less from charging than others, despite having the same band features at higher temperature. This is presumably due to local variations in the residual conductivity, and thus ease of replenishing the photocurrent. We further reduced the beamline flux to a very low value. In this way, we were able to measure the temperature-dependent electronic structure down to $T\approx40$~K with negligible shifting of the bands when doubling the photon flux (indicating an absence of charging). Below 40~K we were not able to fully avoid sample charging effects, but in the worst case, in our data taken at $T$ = 25~K and $h\nu$ = 61~eV, only a moderate rigid shift of the bands ($\approx{55}$~meV) was observed when doubling the flux, without significant broadening of the features of interest. By analysing only the lowest-flux data and also consistently referencing the energy scale to the valence band maximum, the sample charging effects are mitigated in the data presented here. 

X-ray absorption spectroscopy (XAS) and x-ray magnetic circular dichroism (XMCD) measurements were performed in the high-field magnet end-station on beamline I10 at Diamond Light Source. The  sample was cleaved \textit{in situ} in the preparation chamber and transferred under ultrahigh vacuum to the high-field magnet chamber. Spectra were measured with 100\% left- and right-circularly polarized light in total-electron yield detection, which probes the top 3-5 nm of the sample \cite{LaanFigueroa2014Review}. The XMCD measured as a function of field showed that at 20 K the sample magnetization saturates at $\sim$0.5 T. A field of 2 T was used for the measurements shown here. A reference sample of Cr$_2$O$_3$ was measured to provide an energy calibration, as well as to confirm that the measured CrGeTe$_3$ is free from surface oxidation.

Atomic multiplet theory calculations were employed to calculate the Cr $L_{2,3}$ XAS and XMCD spectra using the electric-dipole transitions $3d^n \to 2p^53d^{n+1}$, where the spin-orbit and electrostatic interactions are treated on an equal footing \cite{Thole1985PRB,vanderLaan2006}. The wave functions of the initial- and final-state configurations are calculated in intermediate coupling using Cowan's atomic Hartree-Fock (HF) code with relativistic corrections \cite{vanderLaan1991PRB,CowanBook}. The atomic electrostatic interactions include the $3d$-$3d$ and $2p$-$3d$ Coulomb and exchange interactions, which are reduced to 80\% and 75\%, respectively, of their atomic HF value to account for the intra-atomic screening \cite{Thole1985PRB}. Hybridization effects are included by  mixing the Cr $3d^3$ and $3d^4\underline{L}$ wave functions, where $\underline{L}$ represents a hole on the neighboring atoms in states of appropriate symmetry. The charge-transfer energies in the initial- and final-state are $E(3d^4 \underline{L}) - E(3d^3)$ = $-1$~eV and $E(2p^5 3d^5 \underline{L}) - E(2p^5 3d^4)$ = $-3$ eV, respectively. An octahedral crystal-field of $10Dq$ = 1.8~eV was included. The hybridization parameter $T = \langle \psi(d^3) | H | \psi(d^4 \underline{L}) \rangle$ was 1.7 eV.  The calculated Cr $L_3$ ($L_2$) line spectra are convoluted by a Lorentzian with a half-width of $\Gamma$ = 0.2 eV (0.4 eV) for the intrinsic lifetime broadening and a Gaussian with a standard deviation of $\sigma$ = 0.15 eV for the instrumental broadening.

\section{Results}

Our ARPES measurements within the paramagnetic state above $T_\mathrm{C}$ are shown in Fig.~\ref{fig1}. Consistent with previous studies~\cite{Zhang2019CrSiTe3ARPES,Suzuki2019PRB,Li2018PRB}, these indicate that the uppermost valence bands are comprised of a set of highly-dispersive states, derived from the Te $5p$ orbitals (Fig.~\ref{fig1}(c-f)). Very similar to the 1T transition metal-dichalcogenides \cite{Watson2019PRL}, the $5p_{x,y}$ orbitals form a pair of quasi-2D states at the valence band top, which are separated by a strong spin-orbit splitting of $\approx500$~meV (Fig.~\ref{fig1}(f)). Additionally, there is a $5p_z$-derived state which has a strong out-of-plane dispersion, reaching a maximum at the bulk $\Gamma$ point (Fig.~\ref{fig1}(e)). 

\begin{figure}
	\centering
	\includegraphics[width=\columnwidth]{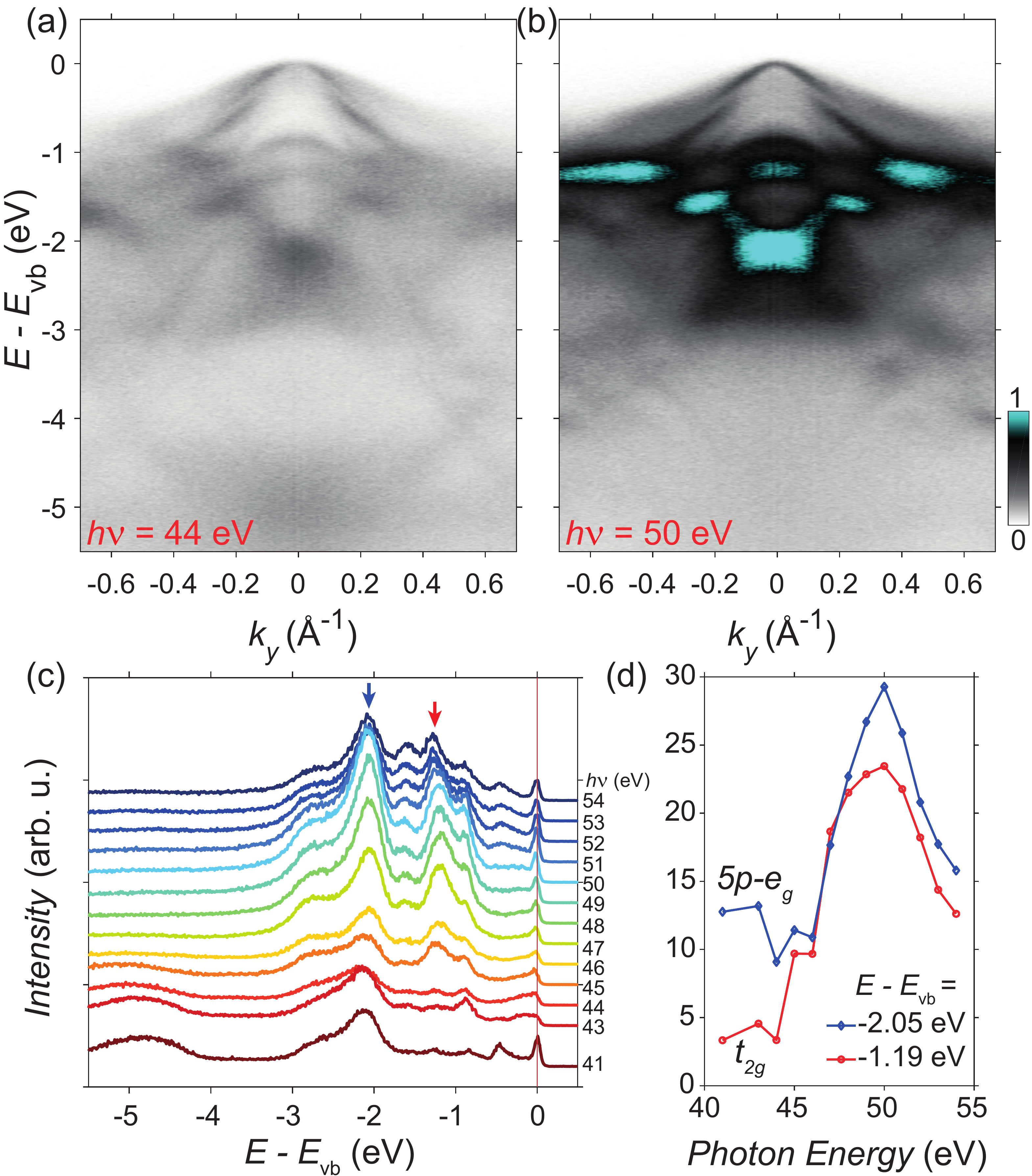}
	\caption{{Resonant ARPES measurements.} a,b) ARPES measurements in the paramagnetic state ($T=80$~K) a) off- ($h\nu=44$~eV) and b) on- ($h\nu=50$~eV) resonance with the Cr $3p$ to $3d$ transition. The colour map is fixed to the same intensity range in both plots, highlighting the extra intensity from the Cr $3d$ weight on resonance. c) Photon energy-dependent energy distribution curves (EDCs) measured at $\overline{\Gamma}$ as a function of photon energy across the resonance. d) The photon energy-dependent spectral weight of peaks located 1.19~eV and 2.05~eV below the valence band top (arrows in (c)), identify the near pure Cr ($t_{2g}$) and partial Cr ($e_g$) character of the corresponding states, indicating that the latter are hybridised with Te $5p$ states.} 
	\label{fig:fig2}
\end{figure}

Here, we focus on the deeper-lying states. Around 0.8-1.5 eV below the top of the valence bands we observe a set of flat bands (Fig.~\ref{fig1}(d)). Their spectral weight is particularly pronounced when measured using photon energies which are tuned into resonance with the core Cr $3p$ to $3d$ transition: they are barely visible when measured using off-resonant photon energies just below this $M$-edge transition in Fig.~\ref{fig:fig2}(a), but dominate the spectrum on resonance (Fig.~\ref{fig:fig2}(b)), before returning to a lower intensity for photon energies above the transition (\ref{fig:fig2}(c)). This reflects a resonant  enhancement of photoemission intensity from states with Cr $d$-orbital character~\cite{Shimada1996}. Quantitative analysis of the spectral weight associated with these states (Fig.~\ref{fig:fig2}(d), red points) allows us to identify the flat bands as having almost entirely Cr $3d$ character. 

\begin{figure}
	\centering
	\includegraphics[width=\columnwidth]{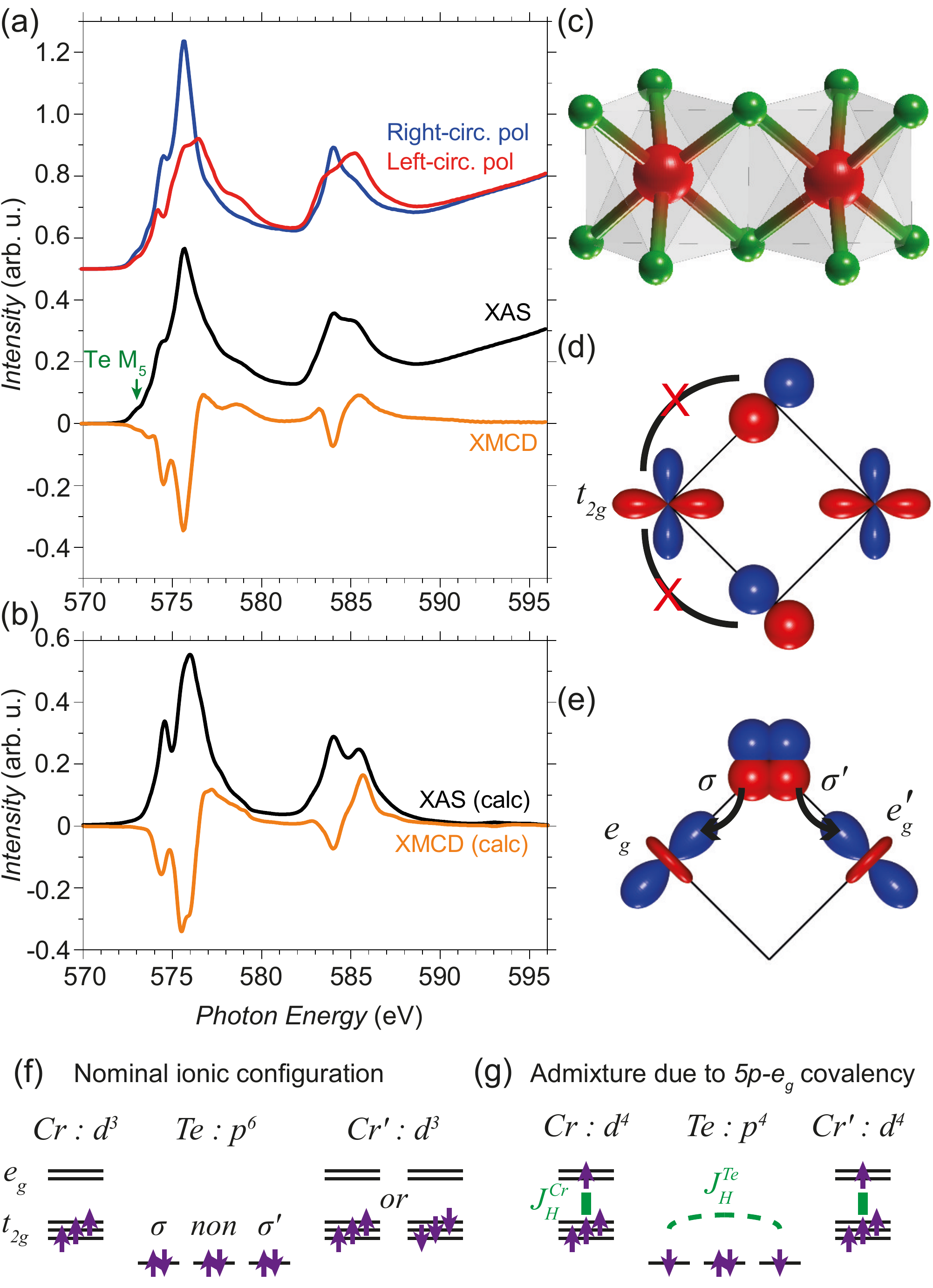}
	\caption{{Evidence for partial occupation of the $e_g$ orbitals from XAS and XMCD}. (a) Dichroic absorption spectra for the Cr L$_{2,3}$ edges measured in an applied field of 2~T which saturates the magnetisation (see methods). b) Simulation of XAS and XMCD from multiplet calculations, reproducing the essential features of the experimental spectra. c) Edge-sharing CrTe$_6$ octahedra. d) For perfect octahedra and exact bond angles, $\sigma$-hopping is forbidden from the $5p$ to the $t_{2g}$ states by symmetry. e) In contrast, at least one of the $e_g$ orbitals is available for $\sigma$-hopping. f) Schematic representation of the nominal ionic configuration of two adjacent Cr atoms and one of the two intermediary Te atoms. The orbitals are defined as in (d,e) with $non$ representing an additional non-bonding $5p$ state. g) Corresponding orbital configuration for a state that additionally hosts a significant $5p\sigma-e_g$ covalent bonding. The unpaired spins on the Te site interact according to $J_H^{Te}$, which gives an energy gain for on-site alignment of spins; this process thus mediates a net ferromagnetic superexchange between the Cr sites.	}
	\label{fig:fig3}
\end{figure}

In the simplest picture for Cr$^{3+}$ with a $3d^3$ configuration, and thus a half-filled $t_{2g}$ shell (Fig.~\ref{fig:fig3}(f)), metallic Cr-derived states would be expected at the Fermi level. From our resonant-ARPES measurements, it is clear that the uppermost Cr-derived states, which we thus assign as $t_{2g}$ states\footnote{Multiple bands are expected here as the $t_{2g}$ manifold is split by trigonal distortions, and because there are two inequivalent Cr sites in the unit cell.}, are located around 1~eV below the valence band maximum, $E_{\mathrm{vb}}$. This reflects how on-site Coulomb repulsion terms ensure the $t_{2g}$ shell remains locally spin-polarised even in the paramagnetic phase, although without long-range order of the spin orientation. Nonetheless, the relatively small binding energy of these states, their small but finite dispersion, and the fact that they remain rather sharp and not completely incoherent in our measurements (Fig.~\ref{fig1}(d)), indicates that the effective on-site Coulomb term is somewhat modest in this system. The overall appearance of this $t_{2g}$ manifold is in strong contrast with that observed in, \textit{e.g.}, PdCrO$_2$, where $t_{2g}$ states of Cr$^{3+}$ form a Mott-insulating sub-system and give rise to a very broad band of incoherent excitations at a higher binding energy of 2 eV \cite{Sunko2018arxiv}. Thus, although the local Coulomb repulsion terms are relevant for understanding the physics of CrGeTe$_3$, it does not reach the limit of fully localized and totally incoherent excitations as suggested by recent dynamical mean field theory calculations \cite{Zhang2019CrSiTe3ARPES}.

Our resonant-ARPES measurements also reveal an additional state located at $\approx\!2$~eV below the valence band maximum which exhibits a roughly two-fold spectral weight enhancement at the Cr $M$-edge resonance (blue points in Fig.~\ref{fig:fig2}(d)). While less pronounced than for the resonant enhancement of the $t_{2g}$ states, our measurements still points to a partial Cr-derived character of this higher binding-energy state. We attribute this to the breakdown of an ionic picture for CrGeTe$_3$. The extensive $5p$ orbitals of Te significantly overlap with the Cr $e_g$ orbitals (Fig.~~\ref{fig:fig3}(e)), and can be expected to form strongly covalent bonding and antibonding states \cite{Kang2019JEM}. The bonding states will be fully occupied, while the antibonding states will form the unoccupied conduction band. 

To confirm this, in Fig.~{~\ref{fig:fig3}(a)} we show XAS and XMCD measurements at the Cr $L_{2,3}$ edge. Compared to a reference sample of Cr$_2$O$_3$ (Cr$^{3+}$, $d^3$), we find that the Cr $L_{2,3}$ edge of CrGeTe$_3$ is shifted to lower photon energies (Fig.~SM2 in S.M. \cite{SM}). Qualitatively, this indicates a higher $d$-electron count in CrGeTe$_3$ as compared to Cr$_2$O$_3$. Additional insight can be gained by comparison with atomic multiplet calculations (see Methods). To obtain the good agreement between the measured and calculated XAS shown in Fig.~{~\ref{fig:fig3}(b)} necessitated treating the ground state of Cr as a hybridized state with approximately 50\% $d^3$ and 50\% $d^4$ character, corresponding to a $d$-electron count of $n_d\approx3.5$. This simultaneously reproduces the measured XMCD spectrum, for which we observe a very strong dichroism (Fig.~\ref{fig:fig3}(a)), and for which our calculations yield physically-reasonable values for the Cr ground state orbital moment of $m_L  = -0.022$  $\mu_{\mathrm{B}}$/Cr and effective spin moment $m_S = 2.81$ $\mu_{\mathrm{B}}$/Cr (\textit{i.e.},  0.80  $\mu_{\mathrm{B}}$ per $3d$ electron).

Further evidence for the hybridisation of Cr $d$ states with Te~$5p$ states comes from the observation of an additional peak in the XAS which we attribute to the Te $M_5$ edge. This is challenging to observe in Cr-based materials as the Te $M_5$ edge lies close in energy to the Cr $L_3$ edge, which will have a significantly larger intensity \cite{Duffy2017PRB}. Nonetheless, in both the XAS and XMCD spectra, we find a small peak at 573 eV, which agrees well with the energy at which the Te $M_5$ edge is expected (green arrow in Fig.~\ref{fig:fig3}(a)). The absence of such a feature in XAS measurements of CrI$_3$ \cite{Frisk2018ML,Kim2019PRL} support that the peak observed at 573~eV here is indeed from the Te $M_5$ edge.  This directly indicates the presence of empty Te $5p$ states above the Fermi level, since the electric-dipole transitions from $3d$ initial states are only allowed to $p$ and $f$ final states.

\begin{figure}
	\centering
	\includegraphics[width=0.9\columnwidth]{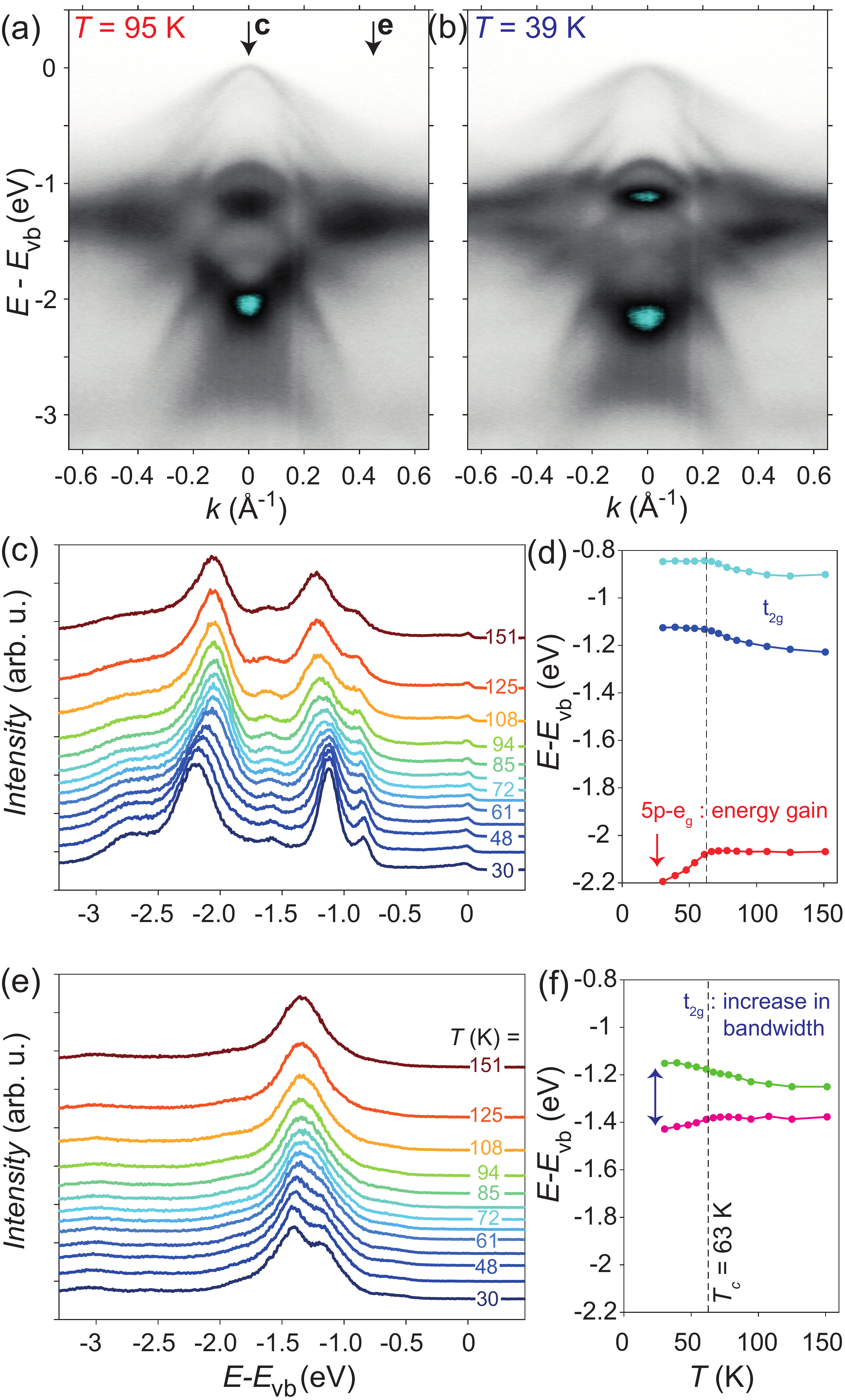}
	\caption{{Temperature-dependent band shifts.} a,b) ARPES measurements through the $\Gamma$ point ($h\nu$~=~48~eV, also on resonance) a) above and b) below $T_\mathrm{C}$. c) Temperature-dependent EDCs at $\Gamma$, and d) corresponding band positions extracted from fitting these EDCs. e,f) Equivalent e) EDCs and f) band positions at $k=0.4$ \AA$^{-1}$ (see arrow in (a)).  }
	\label{fig:fig4}
\end{figure}

On the basis of the XAS, XMCD, and resonant ARPES measurements, we thus attribute the state evident in our ARPES measurements at $\approx\!2$~eV  below the valence band top as a bonding $5p-e_g$ hybridised state. As shown in Fig.~\ref{fig:fig4}(a-d), these $5p-e_g$ states at $\Gamma$ shift to lower energy below $T_\mathrm{C}$. The large magnitude of the shift ($\sim\!130$~meV), and its sharp onset at $T_\mathrm{C}$, are indicators that this state is the leading candidate for driving the ferromagnetic ordering in this system. In contrast, the Cr $t_{2g}$ states exhibit more gradual shifts. These begin at temperatures above $T_\mathrm{C}$, and (from examining the overall spectra, and from analysis of EDCs at momenta away from the high-symmetry point, \textit{e.g.}, Fig.~\ref{fig:fig4}(e,f)) mostly reflect an increase in bandwidth of the $t_{2g}$ states with decreasing temperature, rather than a rigid band shift.

The dichotomy in the temperature-dependent evolution of the $5p-e_g$ and $t_{2g}$ states through $T_\mathrm{C}$ provides important insights into the exchange mechanism underpinning ferromagnetism in CrGeTe$_3$. The long Cr-Cr bond length (3.944 \AA) means that direct exchange will be minimal. Moreover, for the edge-sharing octahedral geometry here (Fig.~\ref{fig:fig3}(c)), with the Cr-Te-Cr bond angle being very close to 90$^\circ$ \cite{Carteaux1995,Ji2013JAPCava}, the formation of $\sigma$-bonds between the $t_{2g}$ orbitals and the anions is strongly suppressed (Fig.~\ref{fig:fig3}(d))~\footnote{For the idealised case, with perfect octahedra and 90$^\circ$ bond angles, the formation of such $\sigma$-bonds is strictly forbidden by symmetry.}. Given this, for the nominal ionic configuration represented in Fig.~{\ref{fig:fig3}(f)}, there would be no mechanism to link the spin directions of the $t_{2g}$ shells of neighbouring Cr sites. However, the $5p-e_g$ hybridisation identified above is fully symmetry-allowed (Fig.~\ref{fig:fig3}(e)). The observation of the corresponding hybridised state in both our ARPES and XAS/XMCD measurements indicates how, beyond the nominal ionic configuration, there is also a significant occupation of orbital configurations with an electron in the $e_g$ shell and a hole in the corresponding $\sigma$ Te orbital (\ref{fig:fig3}(g)).

This provides a natural route to mediate the magnetic interactions via a superexchange mechanism involving the Te atoms. Due to the strong Hund's rule coupling on the Cr sites, the $e_g$ state must be occupied parallel to the local spin orientation of the $t_{2g}$. When two such bonds are formed simultaneously towards the two neighbouring Cr atoms, as shown in Fig.~\ref{fig:fig3}(e,g), the unpaired electrons in orthogonal $\sigma$ and $\sigma'$ orbitals also interact on the Te site, according to the Hund's rule $J^{Te}_H$. This therefore gives an energetic incentive for the development of a net spin moment on the Te sites, antiparallel to the neighbouring Cr sites, and overall mediating a ferromagnetic alignment of spins between the Cr sites. Consistent with this, we note that the Te M$_5$ x-ray absorption edge discussed above (Fig.~\ref{fig:fig3}(a)) exhibits a negative XMCD signal, with the same sign as the Cr $L_3$ edge. Since the azimuthal quantum numbers for the orbitals in these two electric-dipole transitions are opposite (Te $3d \to 5p$ and Cr $2p \to 3d$, respectively), this means that the Te and Cr moments are aligned antiparallel.

The above discussions advance a description of the magnetic ordering in CrGeTe$_3$ via a double exchange mechanism, since both kinetic (delocalisation of $5p$ orbitals) and potential (Hund's coupling on both sites) terms are involved. Indeed, it can be considered as a classic example of how superexchange generally favours ferromagnetism when the relevant bond angles are nearly 90$^\circ$~\cite{Kanamori1959}. As it necessarily involves two separate anion orbitals, it is a weaker interaction than the standard 180$^\circ$ superexchange mechanism involving only one anion orbital, consistent with the much lower transition temperature in CrGeTe$_3$ than \textit{e.g.} antiferromagnetic LaCrO$_3$ ($T_\mathrm{N}$ = 295~K). The key experimental signature validating the dominance of the Te $5p$-mediated exchange mechanism here is the observation of the energy-lowering of the bonding $5p-e_g$ state below $T_\mathrm{C}$ (Fig.~\ref{fig:fig4}(d)), discussed above. This is a signature of the Hund's energy gain on the Te site between pairs of these bonding states, and in fact our measurements here can be taken as one of the most direct confirmations to date of the validity of Kanamori's elegant reasoning regarding 90$^\circ$ superexchange~\cite{Kanamori1959}. 

\begin{figure*}
	\centering
	\includegraphics[width=\textwidth]{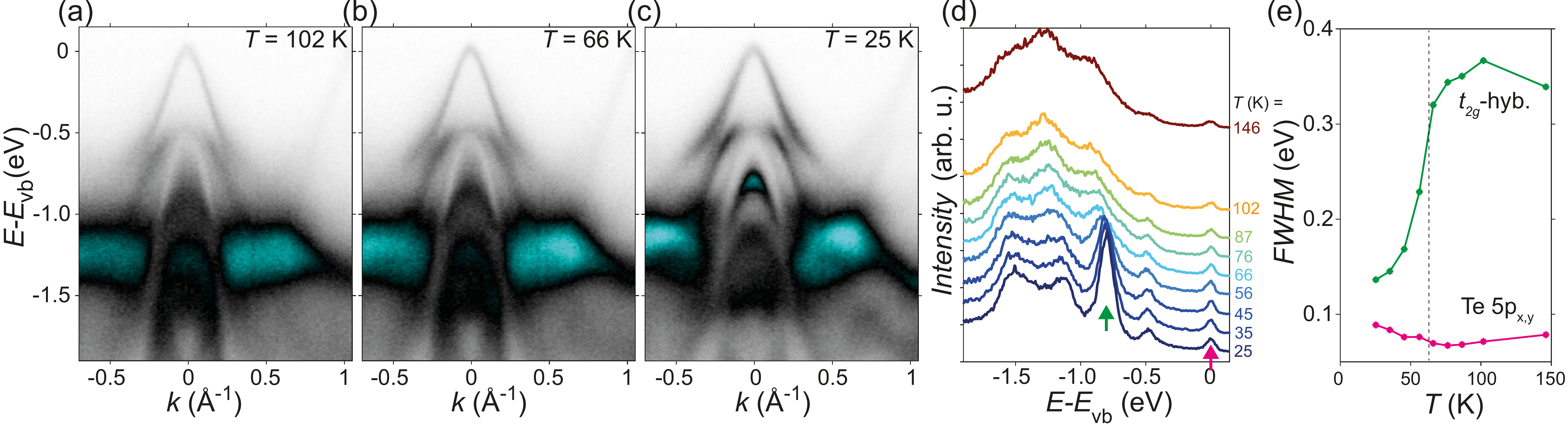}
	\caption{{Lifetime enhancements below $\boldsymbol{T_\mathrm{C}}$.} a-c) Temperature-dependent ARPES measurements at the T point ($h\nu=61$~eV) at temperatures of a) 102~K, b) 66~K, and c) 25~K. d) Corresponding EDCs at the T-point. e) The extracted full-width at half maximum of the uppermost $t_{2g}$ state, determined from fits to these EDCs, shows a remarkable sharpening upon cooling thought the ferromagnetic ordering transition. This indicates a pronounced increase in lifetime of this state, in contrast to the almost temperature-independent lifetime of the uppermost valence band states. For fit details, see Fig.~SM3 in S.M. \cite{SM}.}
	\label{fig:fig5}
\end{figure*}
The resulting emergence of long-range magnetic order below $T_\mathrm{C}$ leaves several other intriguing fingerprints in the spectral function. The first is the increase in bandwidth of the $t_{2g}$ states below $T_\mathrm{C}$ (Fig.~\ref{fig:fig4}(f)). Although $pd\sigma$-hoppings are geometrically suppressed for the $t_{2g}$ states (forbidden in the case of perfect octahedra), direct $dd$ (and some $pd\pi$) hopping processes are still allowed. Thus, in the ferromagnetic ground state, the $t_{2g}$ states hybridise to a limited degree, developing a finite dispersion (Fig.~\ref{fig:fig4}(b)). As discussed above, in the paramagnetic state above $T_\mathrm{C}$, the local $S \approx 3/2$ moment is still formed but fluctuations destroy long-range order. A large energy penalty is thus incurred due to the Hund's interaction if an electron hops onto a neighbouring site with a different spin orientation. The effective $dd$ hopping processes will therefore be suppressed, consequently narrowing the $t_{2g}$ bandwidth. Experimentally, we find that the bandwidth evolves slowly with increasing temperature, and continues to decrease well above $T_\mathrm{C}$ (Fig.~\ref{fig:fig4}(f)). This suggests that the probability of $dd$ hopping processes scales with the spin-spin correlation length, which will be gradually suppressed with increasing temperature above $T_\mathrm{C}$ where short-range spin fluctuations are known to persist~\cite{Williams2015PRB,Zeisner2019PRB,Ron2019NatComms}.

Furthermore, we find that the Cr-derived states measured in ARPES exhibit a particularly-large increase in lifetime upon cooling through $T_\mathrm{C}$. While this is already evident from a pronounced reduction in linewidth of both the Cr $t_{2g}$ and hybridised $5p$-$e_g$ states at the $\Gamma$ point (Fig.~\ref{fig:fig4}), it is particularly apparent for a somewhat more dispersive state 0.8~eV below the valence band maximum at the T point, shown in Fig.~\ref{fig:fig5}. The relatively steep in-plane dispersion of this state likely reflects finite hybridisation with Te $5p$ orbitals, where they become nearly degenerate with the Cr $t_{2g}$ states. While the Te $5p_{x,y}$ states at the valence band top exhibit only weak linewidth changes with increasing temperature, the linewidth of this hybridised Cr-Te state increases by more than a factor of two at the ferromagnetic transition (Fig.~\ref{fig:fig5}(d,e)). Although it is normal for ARPES spectra to display temperature-dependent linewidths due to electron-phonon coupling and other scattering mechanisms, the magnitude of the change here is remarkable. We attribute this to a much reduced probability of spin-flip scattering in the ferromagnetic state, where the spins are aligned, pointing to a significant influence of magnetic fluctuations on shaping the underlying electronic structure of the paramagnetic state. The particular sensitivity of the state identified here likely reflects its more substantial degree of itineracy, and thus a more extensive wavefunction which will exhibit greater sensitivity to longer-range spin alignment. Furthermore, our observations show an interesting correspondence with the substantial increase in phonon lifetimes below $T_\mathrm{C}$ as reported by Raman scattering measurements \cite{Tian2016}, suggesting that electron-phonon as well as spin-phonon scattering could contribute to that effect.  

\section{Conclusions}
Our observations point to a key interplay of itinerant and localized sectors in underpinning the ferromagnetic ordering of CrGeTe$_3$, and in dictating the global electronic structure evolution through $T_\mathrm{C}$. Due to the structural similarity between CrGeTe$_3$ and the CrX$_3$ family, we expect the ideas developed here to be applicable also in the trihalides. Indeed, it is notable that the magnetic ordering temperature in that family increases with increasing covalency: from 16 K in CrCl$_3$ to 61 K in CrI$_3$ \cite{Abramchuk2018,Li2019AdvMat}, while a substantial occupation of Cr $e_g$ orbitals has also been detected in CrI$_3$ \cite{Frisk2018ML}. Excitingly, high-quality ARPES measurements can now be performed on exfoliated flakes of 2D materials~\cite{Cucchi2019,Nguyen2019}. Thus the spectroscopic approach outlined here opens powerful new routes to study how the magnetic ordering tendencies evolve when the material is thinned to the few-layer limit, where $T_\mathrm{C}$ is known to be strongly suppressed~\cite{Gong2017Nature}. We have shown how ARPES measurements provide direct access to the energy scales ultimately underpinning the formation of ferromagnetism; the observation of temperature-dependent band shifts such as those observed here will thus allow determining whether the driving force for ferromagnetism is weakened when interlayer coupling is removed, or whether the suppression of $T_\mathrm{C}$ can instead be attributed to the growth of fluctuations in the 2D limit. More generally, the results shown here indicate how an experimental band-structure perspective can give important insight even in a ``local moment" magnetic system.

\section*{Acknowledgments}

\noindent We thank Cephise Cacho, Chris Hooley, and Shoresh Soltani for useful discussions. We thank Giovanni Vinai and Jun Fuji for support during beamtime measurements. We gratefully acknowledge The Leverhulme Trust (Grant No.~RL-2016-006), The Royal Society, and the European Research Council (Grant No. ERC-714193-QUESTDO) for support. We thank Diamond Light Source for access to beamlines I05 (proposal numbers SI21986, NT22794-3) and I10 (proposal number MM23785) and Elettra synchrotron for access to the APE beamline, which all contributed to the results presented here. This work has been partly performed in the framework of the Nanoscience Foundry and Fine Analysis (NFFA-MIUR, Italy) facility. I.M. and E.A.M acknowledge financial support by the International Max Planck Research School for Chemistry and Physics of Quantum Materials (IMPRS-CPQM).

%

\end{document}